# Cluster dynamics modeling of hydrogen saturation retention in tungsten with a universal trapping-site sink strength

Yuanyuan Zhang[1,2], Xiaoru Chen[1,2] Chuanguo Zhang[1] and Yonggang Li[1,2*]

[1] Key Laboratory of Materials Physics, Institute of Solid State Physics, HFIPS, Chinese Academy of Sciences, Hefei 230031, China
[2] University of Science and Technology of China, Hefei 230026, China
Corresponding author
E-mail: ygli@theory.issp.ac.cn.

## Abstract

Hydrogen isotope (HI) retention poses a key issue for tungsten (W)-based plasma-facing materials (PFMs) in fusion devices, where microstructures such as dislocations (DLs) and grain boundaries (GBs) play a dominant role. Existing theoretical sink strength models for microstructures like DLs and GBs fail to account for the observed saturation of HI retention. In this study, we propose a novel universal trapping-site model that dynamically represents sink strengths as time-dependent site concentrations, which is incorporated into an improved cluster dynamics model for high-fluence HI irradiation. Our simulations quantitatively reproduce the saturated low-energy deuterium (D) retention and depth profiles in W, in good agreement with experiments. A critical saturation fluence of approximately $10^{23}$ m$^{-2}$ is identified, below which unsaturated D retention is governed by both GBs and ion-induced defects, whereas above this threshold GBs dominate D retention by trapping free D and approaching their theoretical saturation limit. The trapping-site sink strength model enables quantification of H trapping by diverse microstructures via unified effective site concentrations, providing mechanistic insights into microstructural effects and facilitating direct evaluation of HI retention in PFMs under different irradiation conditions.

## 1. Introduction

Plasma-facing materials (PFMs) in fusion reactors are subjected to high thermal loads, high-energy neutron irradiation, and high-flux particle impacts generated by D-T reactions, resulting in radiation damage like the formation of basic defects such as vacancies, self-interstitial atoms (SIAs), hydrogen isotope (HI), and helium (He) [1,2]. These defects will evolve into dislocation loops, cavities, and bubbles, or be absorbed by inherent sinks such as dislocations (DLs) and grain boundaries (GBs). In addition, hydrogen (H)/He atoms can easily combine with vacancies to form clusters or bubbles, which are the main sources of crack growth, volume expansion, and radiation hardening in PFMs [3,4]. Tungsten (W)-based materials have been used as PFMs in ITER and considered as a promising candidate for future fusion devices, due to the high thermal conductivity, melting point, and mechanical strength, as well as low sputtering yield, HI, and He retention and erosion rate [5]. But a significant amount of H/He retention usually occurs in W during tokamak operation, mainly due to the influence of surface damage/microstructures on the basic diffusion, trapping and accumulation behaviors of H/He species [6]. This persistent enhancement underscores HI retentions in W-based PFMs as a critical challenge directly governing the safety and operational stability of fusion reactors.

Microstructures in materials significantly affect H/He retention behaviors of W-based PFMs, including impurities (such as C, N and O), second phase particles (oxide/carbide dispersion-strengthened (ODS/CDS) particles), solid solution atoms (such as Ta, Cr, V, Nb, Ti and Re), DLs, GBs, and so on [6–11]. DLs are proven to act as trapping sites, but also as pathways for preferential D diffusion through bulk [12]. The density, distribution and type of DLs further influence these effects by modifying defect sink efficiencies and strain gradient. Similarly, GBs served as the trapping centers can drive H to segregate towards them, leading to a large amount of H accumulation nearby [13,14]. In addition, GBs accelerate H diffusion, promoting rapid H accumulation and retention through enhanced interfacial mobility and localized trapping [15]. Second-phase particles could introduce interfaces and strain fields that act as H trapping sites [16], while solute atoms could alter H solubility and diffusion via lattice distortion and chemical interactions. In summary, the influence of diverse microstructures on HI behavior in W exhibits significant complexity. Inherent defects, particularly DLs and GBs, are the primary candidates responsible for the trapping of H atoms in W under sub-threshold plasma energy exposure conditions [7]. It is necessary to investigate the fundamental mechanisms of HI retention in the inherent microstructures of W-based PFMs.

For experimental researches on H retention in W, the commonly used methods include Ion Beam Analysis (IBA), Secondary Ion Mass Spectrometry (SIMS), Residual Gas Analysis (RGA), Nuclear Reaction Analysis (NRA), Thermal Desorption Spectroscopy (TDS) and so on. These techniques offer valuable insights into the behavior of H in W under various conditions, such as determining H concentration, diffusion characteristics and trapping behaviors. However, each method has inherent limitations. For example, NRA can quantify H concentration but struggles to distinguish between different trapping sites, and TDS can estimate H trapping energies from desorption spectra but cannot directly resolve dynamic desorption pathways or complex defect interactions [3]. Additionally, these techniques are sensitive to experimental conditions like temperature and irradiation dose, making it challenging to infer microscopic mechanisms from static observations [17]. Crucially, the complex, non-equilibrium processes involved in radiation damage, such as defect evolution and clustering, are difficult to trap in real time. These limitations highlight the need for theoretical simulations to complement experimental studies. Computational approaches, especially at the mesoscale, are a preferred alternative approach for providing new insights into H retention mechanisms by modeling defect evolution as well as H diffusion, trapping and desorption.

In theory, Object Kinetic Monte Carlo (OKMC) models describe H interactions with microstructures (such as DLs and GBs) as discrete objects with defined energy states and geometries [18–20]. DLs and GBs are typically idealized as infinite sinks, exemplified by Jansson's cylindrical dislocation model [20] and Valles' perfect-sink GB approximation [21], where all defects reaching these microstructures are permanently removed or absorbed. Due to the spatial correlation and random distribution of defects in the system, OKMC has been commonly adopted to describe the dynamic evolution of complex clusters and microstructures, and quantify H trapping efficiency under non-equilibrium conditions [6]. However, its computational complexity makes it difficult to apply to large-scale and long-time systems. Cluster dynamics (CD) is a crucial tool for studying long-term defect dynamic evolution and H/He retention behavior in materials especially under high-dose irradiation [6]. It is based on the numerical solution of rate equations with the absorption rate of microstructures or sink strength calculated using the steady-state equations proposed by Brailsford and Bullough during the 1970s to 1980s [22–25]. However, these analytical expressions, derived under the mean-field approximation, fail to characterize the dynamic trapping-detrapping behavior of H under non-steady-state conditions. Furthermore, existing OKMC and CD models inherit the critical limitation in the assumption of infinite microstructural absorption capacity, that is, all mobile defects produced by ion irradiation are irreversibly trapped. This neglects saturation and desorption processes, and thus contradicts the experimental observed saturation behavior of H retention. Therefore, it is necessary to develop a universal sink strength model incorporating finite trapping capacities and dynamic defect-microstructure interactions, to accurately describe the saturation and desorption behaviors of H in W with complex microstructures.

It is widely proved by atomic-scale calculations density functional theory (DFT)/molecular dynamics (MD)) that H atoms preferentially occupy discrete lattice sites, particularly tetrahedral interstitial sites (TIS) in W [26–28]. Based on this fact, we propose that the H trapping capacity of microstructures is fundamentally determined by the density of TIS and their interactions with H. By uniformly characterizing microstructural features through TIS concentration and establishing functional relationships between TIS occupancy and defect geometry, the sink strength can be directly linked to the time-dependent site concentration of microstructures. This approach not only eliminates the parametric dependence on complex defect morphologies in traditional analytical models, but also provides a self-consistent description of H retention dynamics via microscopic site trapping. Furthermore, this method could be extended to study the synergistic effects of various types of microstructures, providing a universal formulation of sink strength for multiscale modeling of H interaction with microstructures.

In this paper, we propose a universal trapping-site sink strength model to describe the non-steady state variations of sink strength for microstructures like DLs and GBs. An improved CD model is developed to simulate H retention in W by considering H saturation trapping and emission processes based on the trapping-site sink strength model. The model successfully reproduces the experimental H retention and depth distribution, particularly the saturation phenomenon of H retention. The microscopic mechanisms and H saturation retention are then explored. Finally, we predict the H trapping capabilities of various typical microstructures in W-based PFMs under typical irradiation conditions.

## 2. Theoretical methods

The CD model based on the mean-field rate theory (MFRT) is employed here, which is extensively used to simulate radiation damage, microstructural evolution, and H/He effects during plasma-material interactions. The model can take into account defect generation, diffusion, reactions and trapping processes. Our IRatMat code (CD-pre) developed based on this scheme has successfully investigated the H/He retention in polycrystalline W/Be under different irradiation conditions, including H/He or H/He-neutron synergistic irradiation with different incident ion energies, fluxes, fluence and temperatures [29–31]. In this study, we developed an improved CD code, IRatMat-Site (CD-site), by proposing a universal trapping-site sink strength model to describe the reaction rates of H with various types of microstructures. In the following, we will briefly introduce the basic framework of IRatMat-Site, focusing on the sink strength theory and the proposed trapping-site sink strength model.

### 2.1 Physical model

In typical CD models, a set of one-dimensional diffusion-reaction equations is employed to describe the evolution of various types of defects, accounting for their diffusion process of mobile defects along the depth and their possible reactions with other defects. The master equation describing the concentration evolution of these defects is given as follows [29,30,32],

$$\frac{\partial C_{\theta}}{\partial t} = G_{\theta} + D_{\theta}\nabla^{2}C_{\theta} + \sum_{\theta'}\left[\omega(\theta',\theta)C_{\theta'} - \omega(\theta,\theta')C_{\theta}\right] - L_{\theta}, \quad (1)$$

where, $C_{\theta}$ is the concentration of defect $\theta$ at a specific depth and time. The basic types of defects $\theta$ for H in W represent self-interstitial ($I$), vacancies ($V$), di-interstitials ($I_2$), hydrogen atoms ($H$) and their complex clusters formed by binary reactions ($I_n$, $V_n$, $HI$ and $H_mV_n$), where $m$ and $n$ are the numbers of defects in a loop/cluster. Here, only $I$, $I_2$, $V$ and $H$ are considered to be mobile while all other defect clusters are considered to be immobile for simplification, given the low concentration of large defect clusters at low incident H energies and room temperature [32].

On the right side of Eq. (1), the first term $G_{\theta}$ (in the unit of $m^{-3} \cdot s^{-1}$) is the production rate. The primary radiation damage for defect generation under H ion implantation is calculated by our open-source Monte Carlo code, IM3D [33]. When W is implanted with high-flux H ions at energies below the displacement threshold, the localized supersaturation of H induces lattice stress, leading to the formation of ion-induced defects within the W matrix [34,35]. Ning et al. developed a numerical methodology to quantify the concentration of Frenkel pairs as ion-induced defects in W and integrated it with CD simulations, successfully replicating the depth-dependent profiles and characteristics of D retention under different irradiation conditions [36]. Here, we employ their framework in conjunction with IM3D to quantify the initial defect population produced by low-energy ion implantation. The second term accounts for the spatial diffusion of mobile defects, $D_{\theta} = D_{\theta}^{0}\exp(-E_{\theta}^{m} / k_{B}T)$ (in the unit of m-

$^{2}\cdot \mathrm{s}^{-1}$) represents the defect diffusion coefficients, where $D_{\theta}^{0}$ is pre-exponential factor, $E_{\theta}^{\mathrm{m}}$ is migration energy, $k_{\mathrm{B}}$ is Boltzmann constant and $T$ is the system temperature respectively. The third term accounts for the cumulative contributions of all forward and reverse reactions, where $\omega(\theta',\theta)$ (in the unit of $\mathrm{s}^{-1}$) is the transition rate coefficient per unit concentration of $\theta'$-type defect clusters transforming into $\theta$-type defect clusters (as listed in Table 1).

**Table 1** Reaction types and corresponding rate coefficients.

| |
|---|
| $I+V \underset{G_{\mathrm{I/V}}}{\overset{k^{+}_{\mathrm{I+V}}}{\rightleftarrows}} 0$ |
| $I+I_{\mathrm{n}} \underset{k^{-}_{\mathrm{I_{n+1}}}}{\overset{k^{+}_{\mathrm{I_{n+1}}}}{\rightleftarrows}} I_{\mathrm{n+1}}$ |
| $I+V_{\mathrm{n}} \xrightarrow{k^{+}_{\mathrm{V_n+I}}} V_{\mathrm{n-1}}(n \geq 2)$ |
| $I+H \underset{k^{-}_{\mathrm{HI}}}{\overset{k^{+}_{\mathrm{H+I}}}{\rightleftarrows}} HI$ |
| $I+H_{\mathrm{m}}V \xrightarrow{k^{+}_{\mathrm{H_mV+I}}} mH$ |
| $I+H_{\mathrm{m}}V_{\mathrm{n}} \xrightarrow{k^{+}_{\mathrm{H_mV_n+I}}} H_{\mathrm{m}}V_{\mathrm{n-1}}(n \geq 2)$ |
| $I_2+I_{\mathrm{n}} \underset{k^{-}_{\mathrm{I_{n+2}}}}{\overset{k^{+}_{\mathrm{I_n+I_2}}}{\rightleftarrows}} I_{\mathrm{n+2}}$ |
| $I_2+V \xrightarrow{k^{+}_{\mathrm{V+I_2}}} I$ |
| $I_2+V_{\mathrm{n}} \xrightarrow{k^{+}_{\mathrm{V_n+I_2}}} V_{\mathrm{n-2}}(n \geq 2)$ |
| $I_2+H_{\mathrm{m}}V_2 \xrightarrow{k^{+}_{\mathrm{H_mV_n+I_2}}} mH$ |
| $I_2+H_{\mathrm{m}}V_{\mathrm{n}} \xrightarrow{k^{+}_{\mathrm{H_mV_n+I_2}}} H_{\mathrm{m}}V_{\mathrm{n-2}}(n \geq 3)$ |
| $V+I_{\mathrm{n}} \underset{k^{-}_{\mathrm{I_{n-1}-V}}}{\overset{k^{+}_{\mathrm{I_n+V}}}{\rightleftarrows}} I_{\mathrm{n-1}}$ |
| $V+V_{\mathrm{n}} \underset{k^{-}_{\mathrm{V_{n+1}}}}{\overset{k^{+}_{\mathrm{V_n+V}}}{\rightleftarrows}} V_{\mathrm{n+1}}$ |
| $V+HI \xrightarrow{k^{+}_{\mathrm{HI+V}}} H$ |
| $V+H_{\mathrm{m}}V_{\mathrm{n}} \underset{k^{-}_{\mathrm{H_mV_{n+1}}}}{\overset{k^{+}_{\mathrm{H_mV_n+V}}}{\rightleftarrows}} H_{\mathrm{m}}V_{\mathrm{n+1}}$ |
| $H+V_{\mathrm{n}} \underset{k^{-}_{\mathrm{HV_n-H}}}{\overset{k^{+}_{\mathrm{V_n+H}}}{\rightleftarrows}} HV_{\mathrm{n}}$ |
| $H+H_{\mathrm{m}}V_{\mathrm{n}} \underset{k^{-}_{\mathrm{H_{m+1}V_n}}}{\overset{k^{+}_{\mathrm{H_mV_n+H}}}{\rightleftarrows}} H_{\mathrm{m+1}}V_{\mathrm{n}}$ |
| $\theta+DL \xrightarrow{k^{+}_{\mathrm{dl-\theta}}} D\theta(\theta=I,I_2,V)$ |
| $\theta+GB \xrightarrow{k^{+}_{\mathrm{gb-\theta}}} G\theta(\theta=I,I_2,V)$ |
| $H+DL \underset{k^{-}_{\mathrm{dl-H}}}{\overset{k^{+}_{\mathrm{dl-H}}}{\rightleftarrows}} DH$ |
| $H+GB \underset{k^{-}_{\mathrm{gb-H}}}{\overset{k^{+}_{\mathrm{gb-H}}}{\rightleftarrows}} GH$ |

The last term $L_\theta$ (in the unit of $\mathrm{m^{-3}\cdot s^{-1}}$) denotes the absorption rate of point defects $\theta$ ($I$, $I_2$, $V$ and $H$) by inherent sinks including DLs and GBs here. Most CD models based on the MRFT typically treat it as follows,

$$L_\theta = k^{+}_{\text{S-}\theta} D_\theta C_\theta = \left(k^{+}_{\text{dl-}\theta} + k^{+}_{\text{gb-}\theta}\right) D_\theta C_\theta , \tag{2}$$

which is the addition of various sinks to defect absorption when considering the existing multiple types of microstructures in the system. People have used it to study the absorption behavior of defects in various materials of fission/fusion reactors such as W, stainless steels, zirconium, ferritic alloys, molybdenum and so on [37–40]. Furthermore, some researchers have introduced additional terms in Eq. (2) to improve the accuracy and reality of simulation results, such as considering temperature effects on H emission and transforming H linear concentrations to volume concentrations in DLs and GBs [32]. Ahlgren et al. [41] expressed the dissociation of mobile defects from sinks following the Arrhenius law that,

$$L_\theta = k^{+}_{\text{S-}\theta} D_\theta \left[ C_\theta - C_{\text{S-}\theta} \exp\left( -\frac{E^{\text{b}}_{\text{S-}\theta}}{k_{\text{B}} T} \right) \right]. \tag{3}$$

The stochastic cluster dynamics model developed by Jourdan et al. [42] further takes into account the unit forms of DLs and GBs densities, and makes the following modifications to $L_\theta$ that [32],

$$L_\theta = k^{+}_{\text{S-}\theta} D_\theta \left[ C_\theta - \frac{b^{\text{m}}}{V_{\text{at}}} C_{\text{S-}\theta} \exp\left( -\frac{E^{\text{b}}_{\text{S-}\theta}}{k_{\text{B}} T} \right) \right], \tag{4}$$

where $S$ is sink type ($m = 1,\ 2$ for DLs or GBs), $k^{+}_{\text{S-}\theta}$ is the sink strength of the microstructure $S$ to the mobile defect $\theta$. $V_{\text{at}}$ represents the atomic volume, where $V_{\text{at}} = a_0^3/2$ for bcc system. $b$ is the Burgers vector. $C_{\text{S-}\theta}$ is the linear concentration of the defect $\theta$ in the sink $S$, and $E^{\text{b}}_{\text{S-}\theta}$ is the binding energy between $S$ and $\theta$.

## 2.2 Trapping-site sink strength model for H absorption and emission

$k^{+}_{\text{S-}\theta}$ in Eqs. (2)-(4) is the sink strength of the microstructure $S$ on the mobile defect

$\theta$ , which represents the absorption rate of sink on defects. Taking inherent microstructures of DLs and GBs as example, the model of $k^{+}_{\text{S-}\theta}$ given by Brailsford et al. around 1980s are mostly used in simulation methodologies like OKMC and CD models. Those model are based on the MFRT, incorporating prescribed boundary conditions and treating defect fluxes within a lossy continuum approximation [22–25]. By solving the steady-state diffusion equation under these conditions, the model yields analytical solutions of sink strength and approximate asymptotic solutions under limiting regimes. Here we take DLs and GBs examples to show the development of the trapping-site sink strength model.

Typically, a dislocation is regarded as a cylindrical lossy effective medium with radius $R_{\text{d}}$ in the cylindrical coordinate system. Considered the only sink for defects in the system, the sink strength is given as follows [22],

$$k^{+}_{\text{dl-}\theta} = Z_{\theta}\rho_{\text{d}}, \tag{5}$$

where $Z_{\theta}$ is the trapping efficiency, which is different for $I$ and $V$ , $Z_{\text{I}}$ and $Z_{\text{V}}$ are typically different because of the difference in their dilatation volumes. $\rho_{\text{d}}$ is the dislocation density.

There are several different expressions for GB models, such as the embedded and the cellular models [23]. The cellular model assumes that polycrystals are single spherical isolated gains without a defined medium around them, with no interaction between particle interfaces. The particles at GBs only originate from the diffusion of defects within the grains, and the conditions for all gains are similar as cells. The expression for the cellular sink strength is given by,

$$k^{+}_{\text{gb-}\theta} = k^{\theta}_{\text{sc}}\left(\frac{\sqrt{k^{\theta}_{\text{sc}}}d}{2}\coth\left(\frac{\sqrt{k^{\theta}_{\text{sc}}}d}{2}\right)-1\right)\times\left(1+\frac{k^{\theta}_{\text{sc}}d^{2}}{12}-\frac{\sqrt{k^{\theta}_{\text{sc}}}d}{2}\coth\left(\frac{\sqrt{k^{\theta}_{\text{sc}}}d}{2}\right)\right)^{-1}, \tag{6}$$

with two limit approximations in extreme cases as,

$$k^{+}_{\text{gb-}\theta} = 6\sqrt{k^{\theta}_{\text{sc}}}/d, \quad \text{when } \sqrt{k^{\theta}_{\text{sc}}}d \to \infty, \tag{7}$$

and

$$k^{+}_{\text{gb-}\theta} = 60/d^2, \qquad \text{when } \sqrt{k^{\theta}_{\text{sc}}}d \to 0. \tag{8}$$

In contrast, the embedded model incorporates a critical formulation where the mean defect loss flux at GB is defined as the mean of the outgoing and incoming fluxes. This approach ensures a physically consistent representation of defect-GB interactions by balancing the net defect annihilation rate that,

$$k^{+}_{\text{gb-}\theta} = \frac{6}{d^2}\left\{3+4\beta-\frac{6\beta(\beta-1)}{\alpha^2} + \left[\left(3+4\beta-\frac{6\beta(\beta-1)}{\alpha^2}\right)^2 - 4\left(\beta^2-\alpha^2\right)\left(2-\frac{3(\beta-1)}{\alpha^2}\right)^2\right]^{1/2}\right\} \times\left(2-\frac{3(\beta-1)}{\alpha^2}\right)^{-2}, \tag{9}$$

with two asymptotic approximations under extreme conditions as,

$$k^{+}_{\text{gb-}\theta} = 6\sqrt{k^{\theta}_{\text{sc}}}/d, \qquad \text{when } \sqrt{k^{\theta}_{\text{sc}}}d \to \infty, \tag{10}$$

and

$$k^{+}_{\text{gb-}\theta} = 57.6/d^2, \qquad \text{when } \sqrt{k^{\theta}_{\text{sc}}}d \to 0. \tag{11}$$

Here, $d$ is gain size, $k^{\theta}_{\text{sc}}$ is the total sink strength from all the "single crystal" microstructures within the grain. $\alpha^2 = k^{\theta}_{\text{sc}}d^2/4$, and $\beta = \alpha\coth\alpha$.

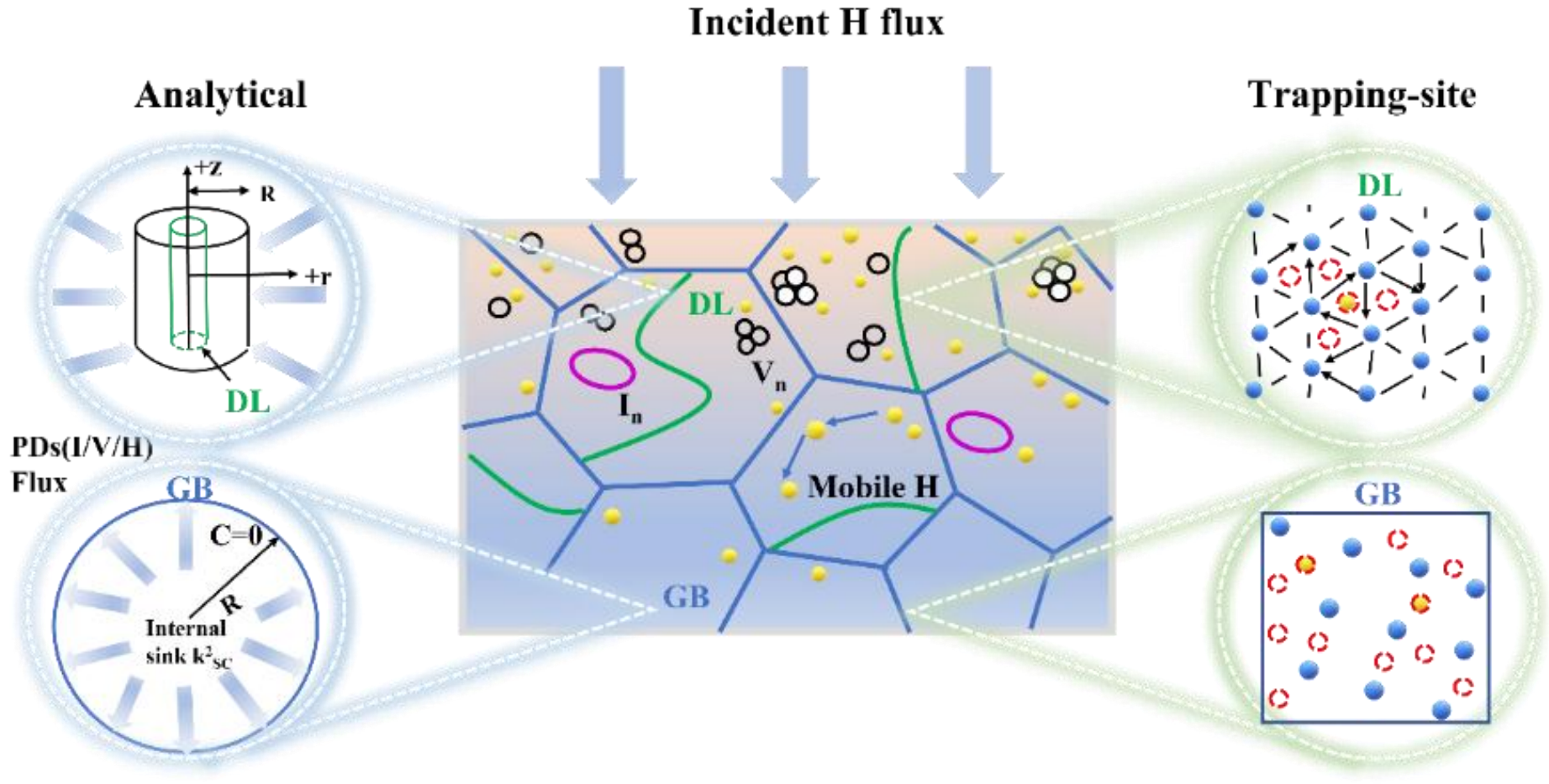

**Fig. 1** Schematic diagram of H diffusion and trapping by various defects and microstructures, based on analytical and trapping-site sink strength models. The blue, red and yellow spheres represent W atoms, TIS and H, respectively.

Those formulations, given by steady-state equation solutions, inherently fail to capture the transient micro-dynamics of sinks. Additionally, the adoption of idealized boundary conditions (Dirichlet assumptions) obscures critical saturation effects in microstructures. Furthermore, the kinetic formulations for DLs and GBs would introduce computational inefficiencies and impede systematic identification of generalizable mechanisms governing H trapping. To break through these limitations, we propose a universal trapping-site sink strength model here to describe both the trapping and emission behaviors of H within microstructures. Due to the relatively small radius and low solubility energy, H tends to occupy the TIS of W in ion channeling experiments [26,43]. This observation is further supported by DFT and MD simulations [44–47]. Liu et al. [45] employed DFT calculations to obtain the formation energies of H at substitutional, TIS, Octahedral interstitial sites and vacancies. They found that the lowest formation energy occurs at a specific TIS. Based on these facts, we propose a fundamental hypothesis that H trapped in sinks is composed of a series of sites, regardless of their structural configurations. H atoms in W generally exhibit a weak repulsion between each other, and exist as single H distributed as evenly as possible on the equivalent sites of microstructures [46,47], which supports the validity of our hypothesis. Therefore, the trapping and emission effects of inherent defects like DLs and GBs on H can be essentially described as the trapping of H by a set of TIS on them. In principle, the interactions between these TIS can be negligible compared to the interaction between H and the TIS, as indicated by atomic simulations [48]. The fourth reaction term on the right side of Eq. (1) can thus be represented instead of Eq. (2) as follows,

$$L_{\text{S-H}} = 2\pi\left(r_{\text{S}} + r_{\text{H}}\right) Z_{\text{S}}^{\text{H}} D_{\text{H}} \rho_{\text{S}} \left\{ n_{\text{S}}(t) C_{\text{H}} - \left[n_{\text{S}}^{0} - n_{\text{S}}(t)\right] \exp\left(-\frac{E_{\text{S-H}}^{\text{b}}}{k_{\text{B}} T}\right) / V_{\text{at}} \right\}, \quad (12)$$

where $2\pi\left(r_{\text{S}} + r_{\text{H}}\right)$ represents the geometric factor, with $r_{\text{S}}$ and $r_{\text{H}}$ are the absorption radius of site $S$ and the radius of $H$ [39]. $Z_{\text{S}}^{\text{H}}$ is a dimensionless factor representing the absorption efficiency of $H$ by the sink $S$, which is usually set as 1.0 [30]. $\rho_{\text{S}}$ is the density of $S$. $n_{\text{S}}^{0}$ is the total number of TIS per unit length/area/volume of zero/one/two dimensional (0-D/1-D/2-D) sinks, which governs the maximum H trapping capacity (or saturation concentration) of $S$. In typical metals, $n_{\text{dl}}^{0}$ and $n_{\text{gb}}^{0}$ are in the order of $10^{10}$ m$^{-1}$and $10^{20}$ m$^{-2}$, respectively. $n_{\text{S}}(t)$ denotes the unoccupied TIS density that varies with time, depth, and sink type. $n_{\text{S}}^{0}-n_{\text{S}}(t)$ indicates the density of TIS occupied by H. Thus, the first and second terms in Eq. (12) correspond to the concentration of H through absorption and thermal emission, respectively. The sink strength of $S$ on $H$ is thus

$$k_{\text{S-H}}^{+}=2\pi(r_{\text{S}}+r_{\text{H}})Z_{\text{S}}^{\text{H}}\rho_{\text{S}}n_{\text{S}}(t) \tag{13}$$

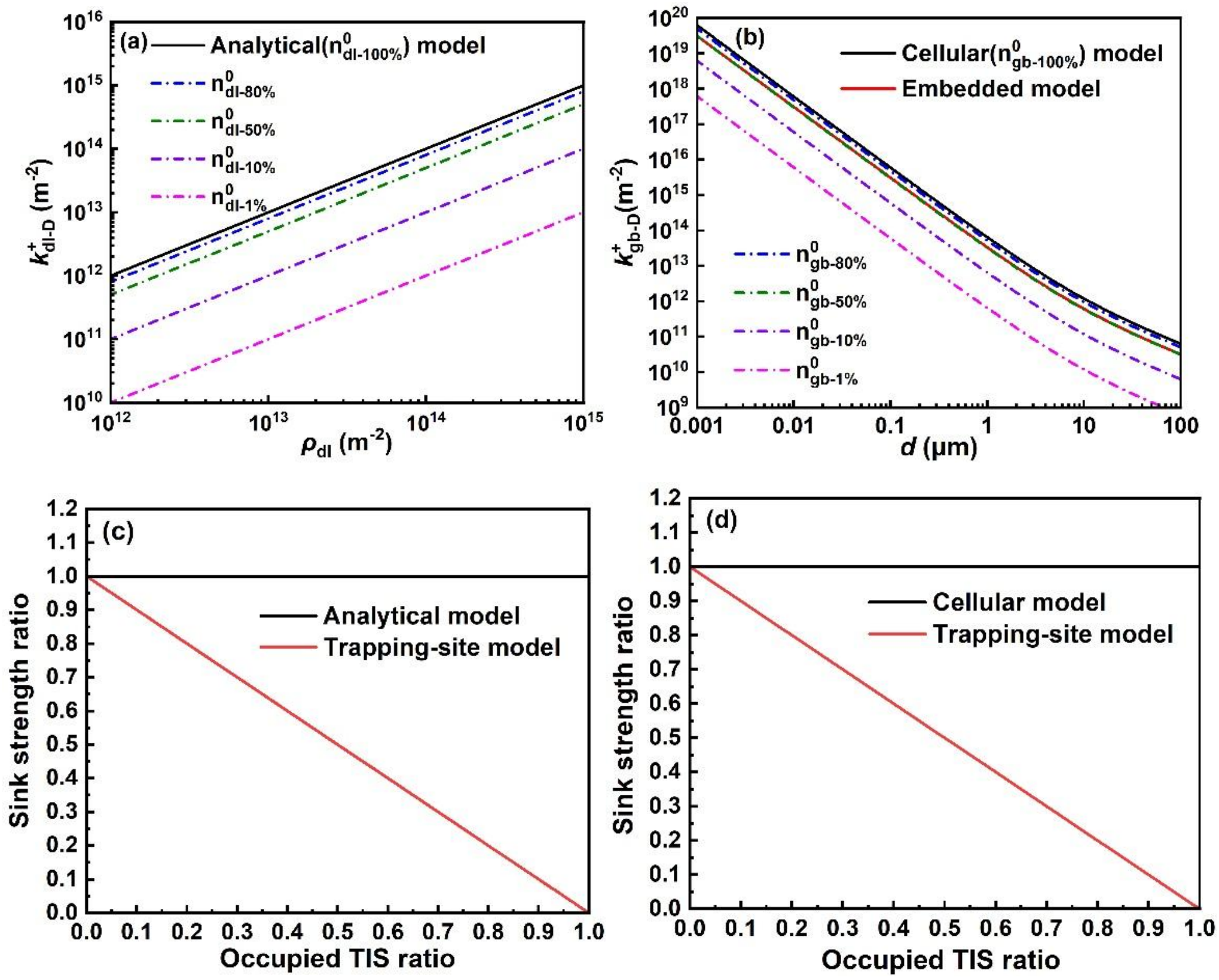

**Fig. 2** Sink strength of the analytical models in Eqs. (5), (8) and (11) as well as the trapping-site sink strength model in Eq. (13) with (a) the dislocation density $\rho_{\text{dl}}$ and (b) grain size $d$, where dashed lines represent different empty site ratios in the trapping-site sink strength model. Sink strength ratio versus occupied TIS ratio for (c) DLs and (d) GBs.

Fig. 2(a) and (b) show the typical relationships of sink strength $k_{\text{dl-D}}^{+}$ and $k_{\text{gb-D}}^{+}$

with dislocation density $\rho_{\text{dl}}$ and grain size $d$, respectively, for the traditional analytical models and the trapping-site sink strength model with different empty site ratios. Fig. 2(a) shows that $k^{+}_{\text{dl-D}}$ increases with increasing $\rho_{\text{dl}}$, which is consistent with Eqs. (5) and (13), indicating a proportional relationship. Fig. 2(b) reveals that $k^{+}_{\text{gb-D}}$ decreases with increasing $d$, due to the reduced grain boundary density. The dashed lines in Fig. 2(a) and (b) show that sink strength declines with the empty site ratio, as inadequate empty sites suppress the trapping capacity of H in sinks. Fig. 2(c) and (d) further confirm that sink strength ratio (the ratio of trapping-site to analytical sink strength) decreases linearly with increasing of occupied TIS ratio ($\left[n_{\text{S}}^{0}-n_{\text{S}}\left(t\right)\right]/n_{\text{S}}^{0}$). Whereas, the analytical model (black solid line) shows sink strength remains constant regardless of occupied TIS ratios. When the occupied ratio is zero ($n_{\text{S}}(t)=0$), the sink strength for the trapping-site model approaches the analytical model, indicating that the latter is the limit unoccupied condition of the former. The trapping-site model characterizes the dynamic trapping of H by sinks, offering a comprehensive description of variations in sink strength.

Compared to the analytical model for sink strength, the trapping-site model offers several advantages. Firstly, the trapping-site model can describe the behavior of H absorbed by microstructures under non-steady state conditions by introducing a time-dependent parameter for site density. Secondly, the finite site density effectively explains the gradual saturation of H retention under high-fluence H ion irradiation. Finally, the trapping-site model is more universal and applicable to various sinks such as voids, impurities, ODS/CDS, DLs, GBs and others. The geometric structure, site density, interaction energy, and other characteristics of these sinks are all incorporated into Eq. (13), which further simplifies the CD calculation process and its implementation.

# 3. Results and Discussion

## 3.1 Model parameters and verification

The reliability of the CD-site model is verified by comparing with experimental results directly [49]. To ensure the reliability and accuracy of our model, we carefully choose the fundamental kinetic and energetic parameters by considering the published values from DFT/MD calculations or experiments, as shown in Table 2. The binding energies of H with various types of DLs and GBs typically exhibit a Gaussian distribution [32]. For simplicity, their mean values are adopted in this study, as our primary focus is on the effects of the newly proposed trapping-site model. In fact, the multi-trapping effects of H at GBs can be treated as in our previous work [32]. Specifically, the binding energies of mobile defects ($I$ , $I_2$ and $V$ ) with large defect clusters were estimated using the Capillary Law [31], while the binding energies of $H$ with $H_mV_n$ clusters are estimated through a simple formula $E^{\mathrm{b}}_{\mathrm{H_mV_n \to H}} = 1.04 \times (m/n)^{-0.43}$ [21]. The reaction types for H in W here were chosen the same as our previous research [32], which has reasonably described the behavior of H retention and desorption in W under deuterium (D) irradiation. And the reaction rate coefficients of H in W are comprehensively illustrated elsewhere [29,31,36,50]. Fig. 3 shows the initial depth distribution of D and D-induced defects in W under low-energy D ion implantation as mentioned in Section 2.1. The ion-induced defects are localized within the same near-surface region as D ions.

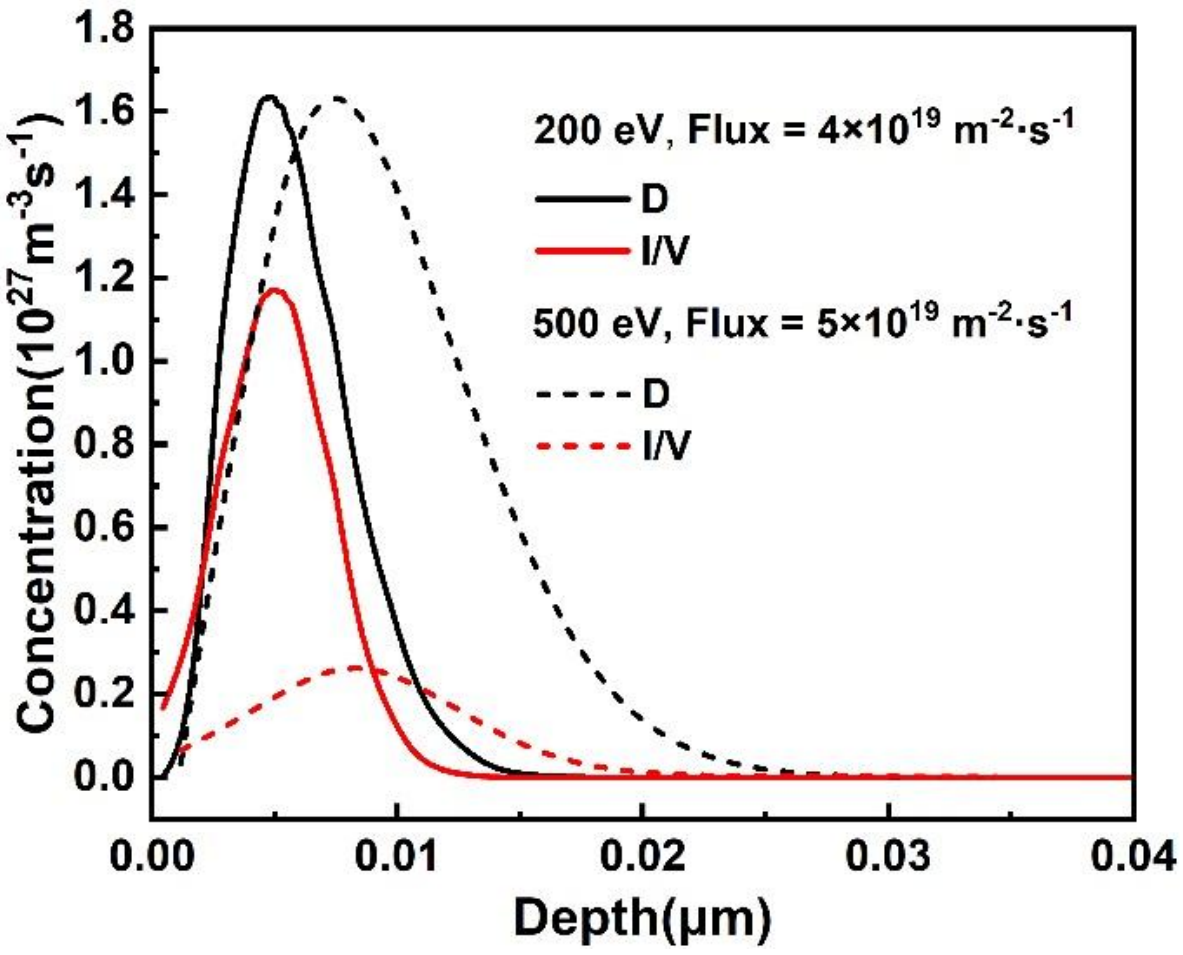

**Fig. 3** Depth distribution of implanted D and D-induced defects $(I/V)$ in W under D ion implantation with 200 eV energy and the flux of $4 \times 10^{19}$ m$^{-2}$ · s$^{-1}$ as well as 500 eV with the flux of $5 \times 10^{19}$ m$^{-2}$ · s$^{-1}$.

**Table 2** Summary of simulation parameters.

| **Parameter** | **Symbol** | **Value** | **References** |
|---|---|---|---|
| Lattice constant | $a_0$ | 3.1652 Å | [51] |
| Recombination radius | $r_{I/V}$ | 7.9 Å | [39] |
| D radius | $r_D$ | 0.53 Å | [52] |
| Reaction radius | $r_S$ | 1.94 Å | [48] |
| SIA/dislocation elastic interaction factor | $Z^I$ | 1.2 | [39] |
| Dislocation line density | $\rho_{dl}$ | $10^{12}$ m$^{-2}$ | [30] |
| SIA pre-exponential factor | $D_{I_0}$ | $8.77\times10^{-8}$ m$^2\cdot$s$^{-1}$ | [53] |
| V pre-exponential factor | $D_{V_0}$ | $1.77\times10^{-6}$ m$^2\cdot$s$^{-1}$ | [53] |
| H pre-exponential factor | $D_{H_0}$ | $4.1\times10^{-7}$ m$^2\cdot$s$^{-1}$ | [54] |
| Migration energy of SIA | $E_I^m$ | 0.013 eV | [55] |
| Migration energy of V | $E_V^m$ | 1.66 eV | [21] |
| Migration energy of H | $E_H^m$ | 0.205 eV | [21] |
| Formation energy of SIA | $E_I^f$ | 9.466 eV | [56] |
| Formation energy of V | $E_V^f$ | 3.80 eV | [56] |
| Formation energy of H | $E_H^f$ | 2.45 eV | [45] |
| Binding energy of $I_2$ | $E_{I_2}^b$ | 2.12 eV | [53] |
| Binding energy of $V_2$ | $E_{V_2}^b$ | -0.115 eV | [57] |
| Binding energy of H-I | $E_{H-I}^b$ | 0.33 eV | [21] |
| Binding energy of H-DL | $E_{dl-H}^b$ | 0.1-1.6 eV | [12,58] |
| Binding energy of H-GB | $E_{gb-H}^b$ | 0.4-1.2 eV | [15,30] |

As shown in Fig. 4, the D retention and depth distribution profiles in W were simulated, for 500 eV D ions implantation with the flux of $5\times10^{19}$ m$^{-2}\cdot$s$^{-1}$ and the fluence ranging from $10^{19}$ to $10^{25}$ m$^{-2}$ at 300 K (Fig. 4(a)) [59–61], and for 200 eV implantation with the flux of $4\times10^{19}$ m$^{-2}\cdot$s$^{-1}$ up to the fluence of $3.2\times10^{24}$ m$^{-2}$ at 323 K (Fig. 4(b)) [49]. Overall, in Fig. 4(a), the CD-site simulation results exhibit a gradual saturation with increasing fluence, consistent with experimental observations. This contrasts with the CD-pre curve, following a linear increase with increasing D fluence. In Fig. 4(b), the CD-site results are also consistent with experimental data compared to the CD-pre results, particularly in the following aspects: 1) a peak in the near surface layer (0-0.2 μm) is on the same order of magnitude as the experimental value; 2) the plateau region (0.7-7 μm), within one order of magnitude of deviation from experimental values and in contrast with the continuous decline with depth in the CD-pre profile. This retention is mainly due to D being trapped by inherent defects in the bulk [62]. We considered two types of inherent defects (DLs and GBs) by analyzing sites in them, leading to a prediction of H saturation retention. At low fluence, D retention exhibits near-linear fluence dependence, due to efficient D trapping by abundant unoccupied sites. At intermediate fluence, the growth becomes sublinear as sites are gradually occupied, leading to fewer available sites and a decline in trapping efficiency. Above the critical saturation fluence of about $10^{23}$ $m^{-2}$, the saturation of sites in DLs and GBs by D shifts the retention into a plateau, where further accumulation becomes negligible.

At the low fluence of $10^{20}$-$10^{22}$ $m^{-2}$ in Fig. 4(a), the simulated D retention is smaller than the experimental values. Both the differences in complex W properties and different H analysis techniques (such as NRA and TDS) could essentially affect the experimental results of D retention more or less [1,3,63,64]. On the other hand, the fundamental energetic parameters, other defects (such as impurities) and the density of DLs/GBs in the model may differ from the experimental conditions. When the incident fluence is low, the influence of these factors on the total D retention is much greater compared to high fluence, that is why our simulation results are noticeably lower than the experimental values. It is worth noting that the D retention experimental data inherently exhibit deviations of about 1-2 orders of magnitude with each other, due to the complex and changeable microstructures and the differences in experimental conditions. Therefore, the discrepancy between the simulation results and the experimental values in this study remains within an acceptable range of variation. The discrepancy between CD-site and experimental data in Fig. 4(b) is the absence of the near-surface concentration dip in the CD-site results. Cavities/blisters will be formed due to the increase of dislocation or void density induced by D supersaturation near the surface under low-energy, high-flux D plasma irradiation [32], which is very complex and should be studied in future research.

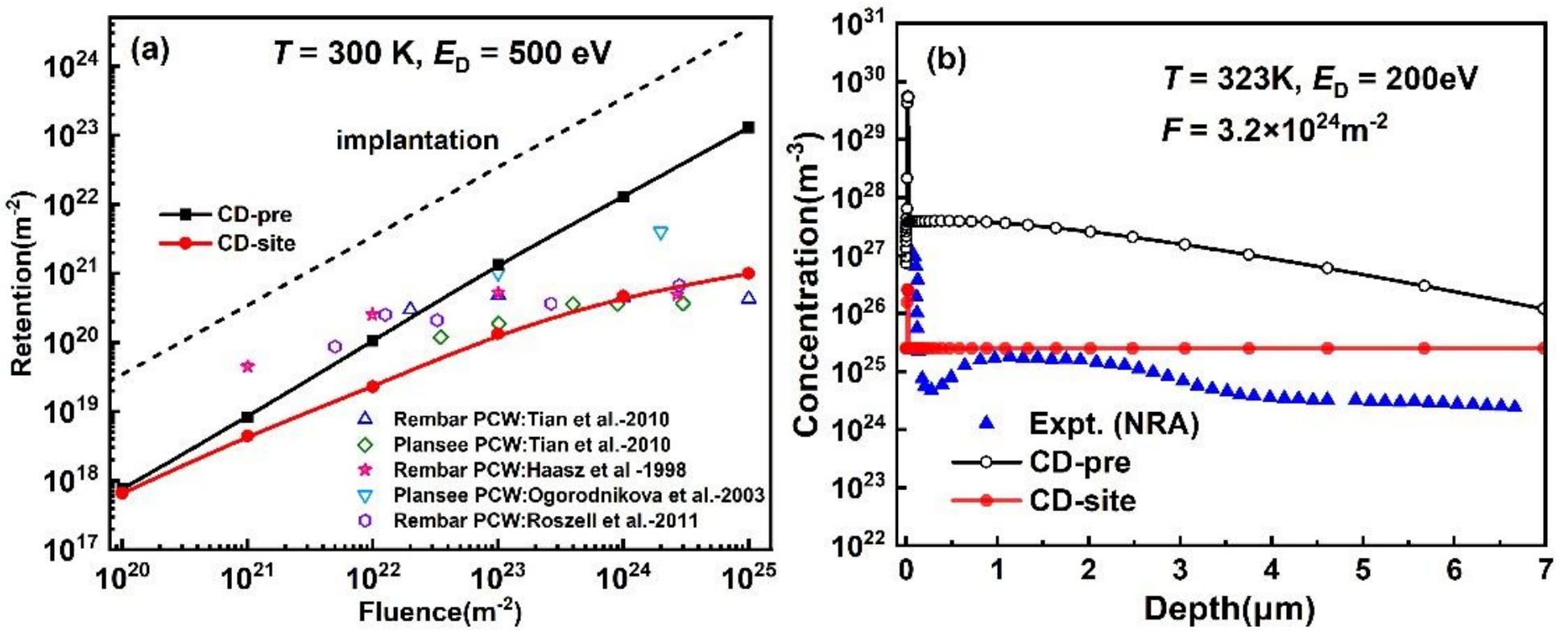

**Fig. 4** Comparison of simulations with experiments. (a) Fluence dependence of D retention in W under 500 eV D implantation at 300 K [59–61]. (b) Depth profiles of D in W after 200 eV D ion implantation with a fluence of $10^{24}$ m$^{-2}$ at 323 K [49].

### 3.2 Effect of various defects/sinks on D retention

Identifying which microstructure plays a key role in H saturation retention is essential for understanding microscopic mechanisms of H evolution behavior in W. Fig. 5 quantifies the retained D in the form of free D and clusters ($DI$ and $D_mV_n$), as well as trapping in DLs, and GBs under different fluences. The CD-site simulation results for DLs and GBs indicate that the retention of D gradually approaches a saturation value as fluence increasing, which is similar to that in Fig. 4(a). In contrast, the CD-pre simulation results show a linear increase in retained D with increasing fluence. The retained D in clusters simulated by both CD-site and CD-pre overlap completely, which also show a saturation trend with increasing fluence when clusters reach their maximum number of trapped D. Thus, we infer that the saturation of D retention mainly originates from the saturation of D trapping in DLs and GBs. Furthermore, since the theoretical limit concentration of D retained in GBs (with a grain size of $10^{-5}$ m) is about $10^{25}$ m$^{-3}$, at least 3 orders of magnitude higher than that in DLs (with a dislocation density of $10^{12}$ m$^{-2}$), the dominant saturation retention for low-energy D arises from trapping in GBs.

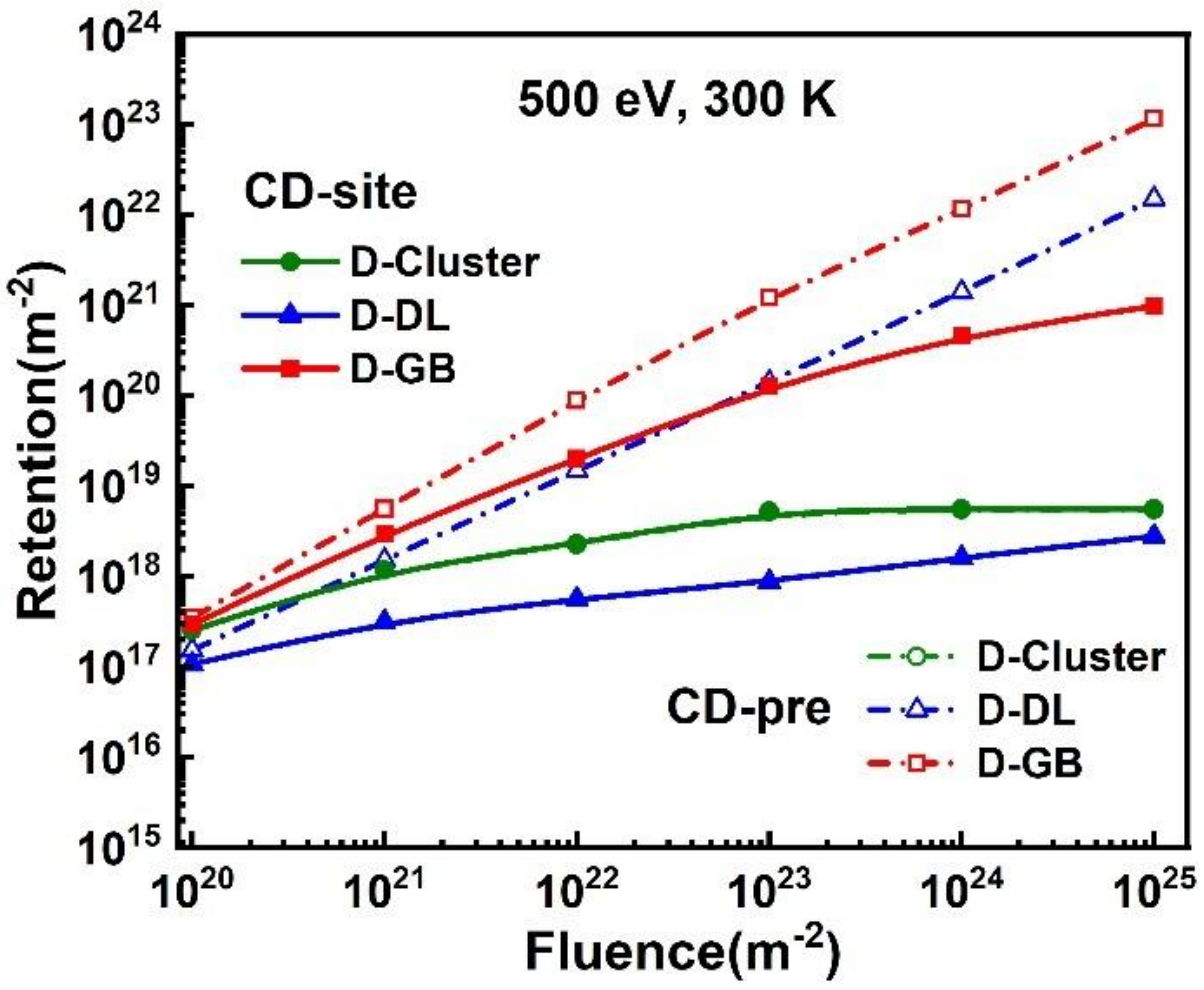

**Fig. 5** D retention in Clusters, DLs and GBs in W under 500 eV D ion implantation with the flux of $5\times10^{19}$ $m^{-2}\cdot s^{-1}$ and different fluences at 300 K, where solid and dash lines represent the CD-site and CD-pre results, respectively.

In order to understand the microscopic mechanisms of D retention in W, the contributions of different forms of D retained in W are shown in Fig. 6. It shows D concentration profiles obtained using CD-pre and CD-site, during D implantation with the fluence of $10^{20}$ and $10^{24}$ $m^{-2}$. The D distribution from CD-pre and CD-site are nearly identical at the fluence of $10^{20}$ $m^{-2}$, while notable differences emerge between them in terms of DLs and GBs at the fluence of $10^{24}$ $m^{-2}$. In particular, the CD-site results show a distinct platform up to the depth of 10 μm. This discrepancy arises because DLs and GBs provide sufficient sites to trap D at low fluence, but as the fluence increases, these sites become progressively saturated with D occupation. In CD-pre, the assumption of infinite D trapping leads to an overestimation of D retention in DLs and GBs. Overall, the retained D in the near-surface region comes from both clusters and GBs. The D concentration peak is primarily dominated by clusters, as explained by a detailed trapping mechanism analysis of D by $D_mV_n$ clusters as shown in Fig. 7. Vacancies, with a high migration energy at room temperature, combine with D in the near-surface irradiated region to form $D_mV_n$ clusters. In contrast, SIAs and H, with lower migration energies, diffuse more deeply. The rapid migration of SIAs promotes I-V recombination in the irradiation region, which in turn decreases the vacancies concentration, causing the $D_mV$ peak to shift backward to 0.015 μm and become narrower. The retained D in the bulk primarily originates from DLs and especially from GBs, as these two types of inherent sinks contain a larger number of sites capable of capturing substantial amounts of D. The finite concentration of sites at GBs imposes a theoretical upper limit on the amount of D that can be retained for CD-site. With increasing fluence, the trapping capability of GBs for D gradually decreases as the limit concentration is approached, allowing more D to diffuse deeper into the bulk. Consequently, the CD-site simulation results exhibit better agreement with experimental measurements compared to those of CD-pre, as shown in Fig. 4(a).

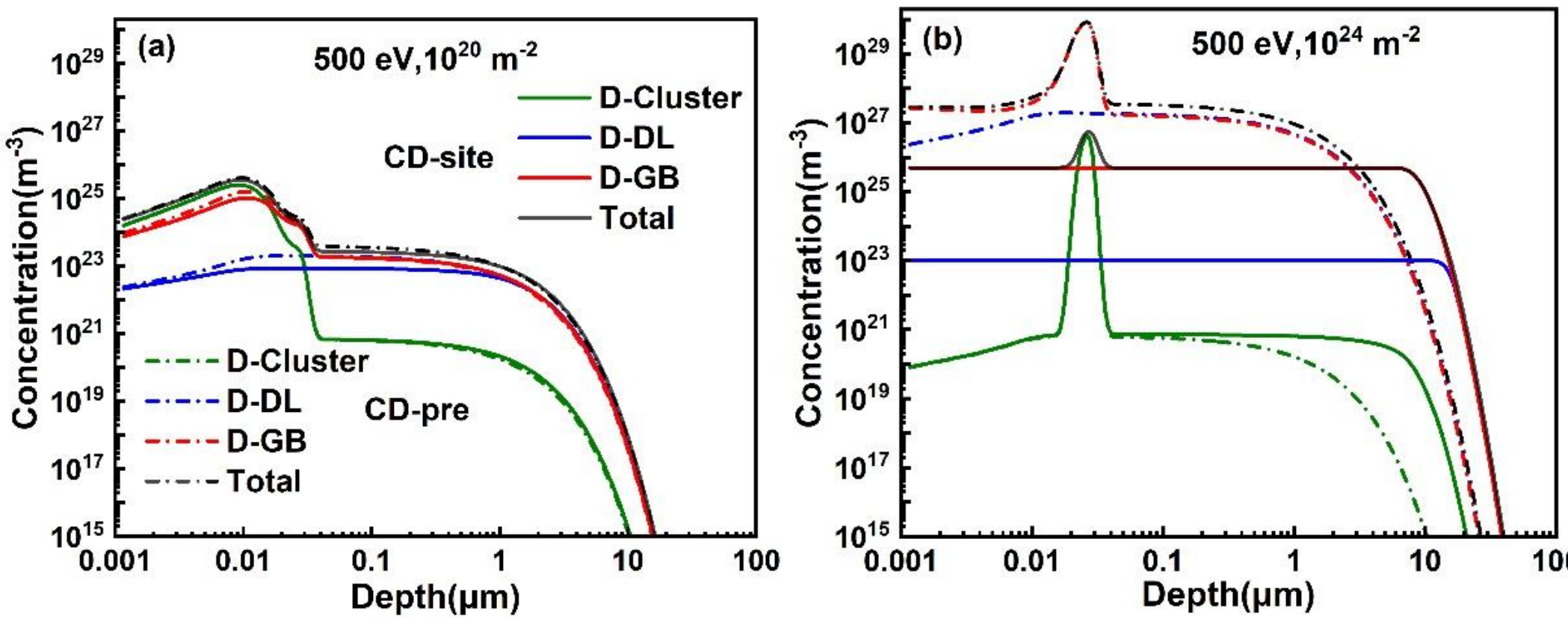

**Fig. 6** Depth distributions of D trapped in clusters (D-Cluster), DLs (D-DL) and GBs (D-GB) during 500 eV D ion implantation with the fluence of (a) $10^{20}$ and (b) $10^{24}$ $m^{-2}$ and the flux of $5\times10^{19}$ $m^{-2}\cdot s^{-1}$ at 300 K.

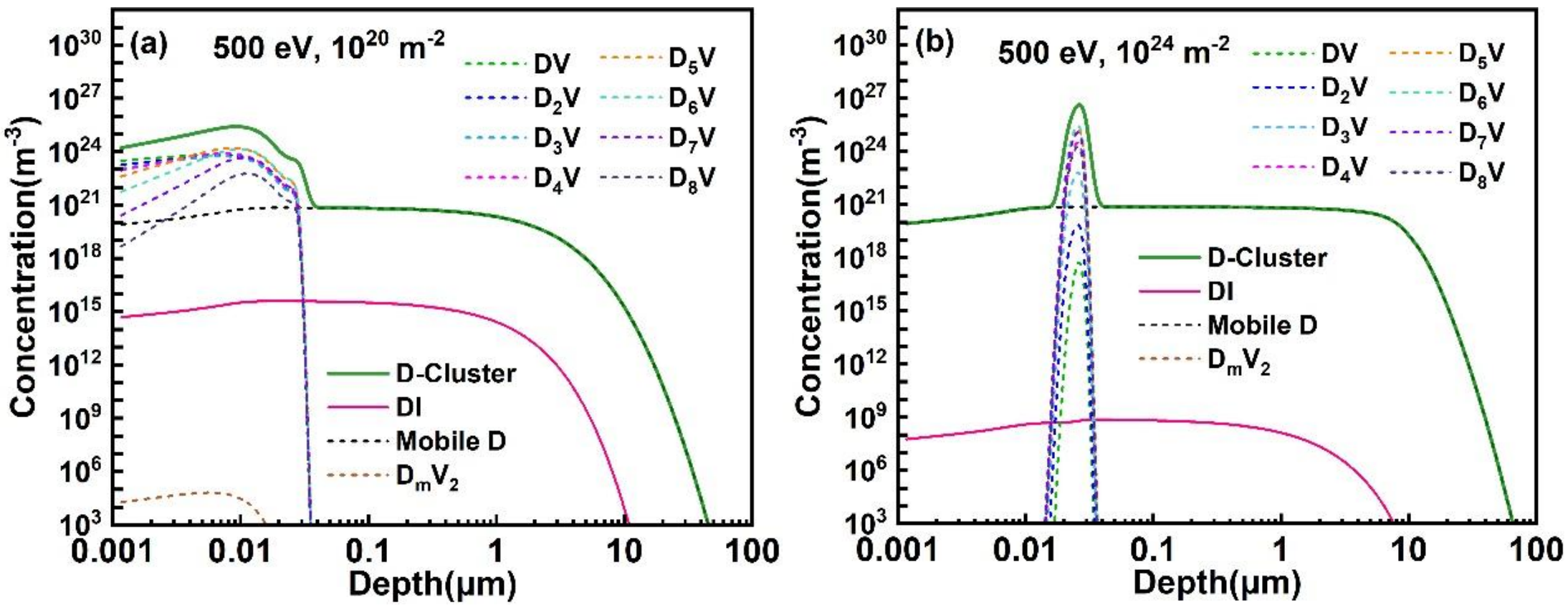

**Fig. 7** Detail depth profiles of D retained in Cluster (D, $DI$ and $D_mV_n$) during 500 eV D implantation with the fluence of (a) $10^{20}$ and (b) $10^{24}$ $m^{-2}$ and the flux of $5\times10^{19}$ $m^{-2}\cdot s^{-1}$ at 300 K.

### 3.3 Distribution and evolution of trapping sites

The above results indicate that inherent microstructures including DLs and GBs dominate D saturation retention under D ion implantation with the fluence higher than $10^{23}$ m$^{-2}$. Within the CD framework, DLs and GBs are typically assumed as spatially uniform distributions. Accordingly, the ratio of occupied site density ($n^0_{\mathrm{dl/gb}} - n_{\mathrm{dl/gb}}(t)$) to total site density ($n^0_{\mathrm{dl/gb}}$) can serve as a key metric for evaluating the D-trapping capacity of inherent microstructures along with depth in materials under different irradiation conditions. Fig. 8 shows the depth and fluence dependence of the occupied sites density in percent $1\text{-}n_{\mathrm{dl/gb}}(t)/n^0_{\mathrm{dl/gb}}$. The black line (with the occupied site density of 95% in percent) divides the distribution into two regions, that is, the unoccupied site region and the saturated region where all sites are fully occupied by D. As noted in Section 3.2, the theoretical limit concentration of D retained in DLs (on the order of $10^{22}$ m$^{-3}$) is much less than that of GBs (on the order of $10^{25}$ m$^{-3}$), causing DLs reach saturation at much lower fluence threshold (less than $10^{20}$ m$^{-2}$) and exhibit a broader saturation region (Fig. 8(a)). As shown in Fig. 8(b), the density of occupied sites at GBs with ion fluence can be divided into three distinct regions. In Region I (unsaturated), the total sites concentration exceeds the implanted D concentration, resulting in over 90% of sites remaining unoccupied and thus minimal D retention. In Region II, D saturation occurs primarily in the near-surface region (within 0.03 μm). In this region, ion-induced defects trap part of the incoming D, preventing the saturation region from extending deeper into W. In region III, the fluence exceeds the critical saturation fluence of $10^{23}$ m$^{-2}$, thus the saturated area rapidly expands. In this interval, all sites in the irradiated region are occupied, and more D diffuse into the bulk and are trapped by the GBs.

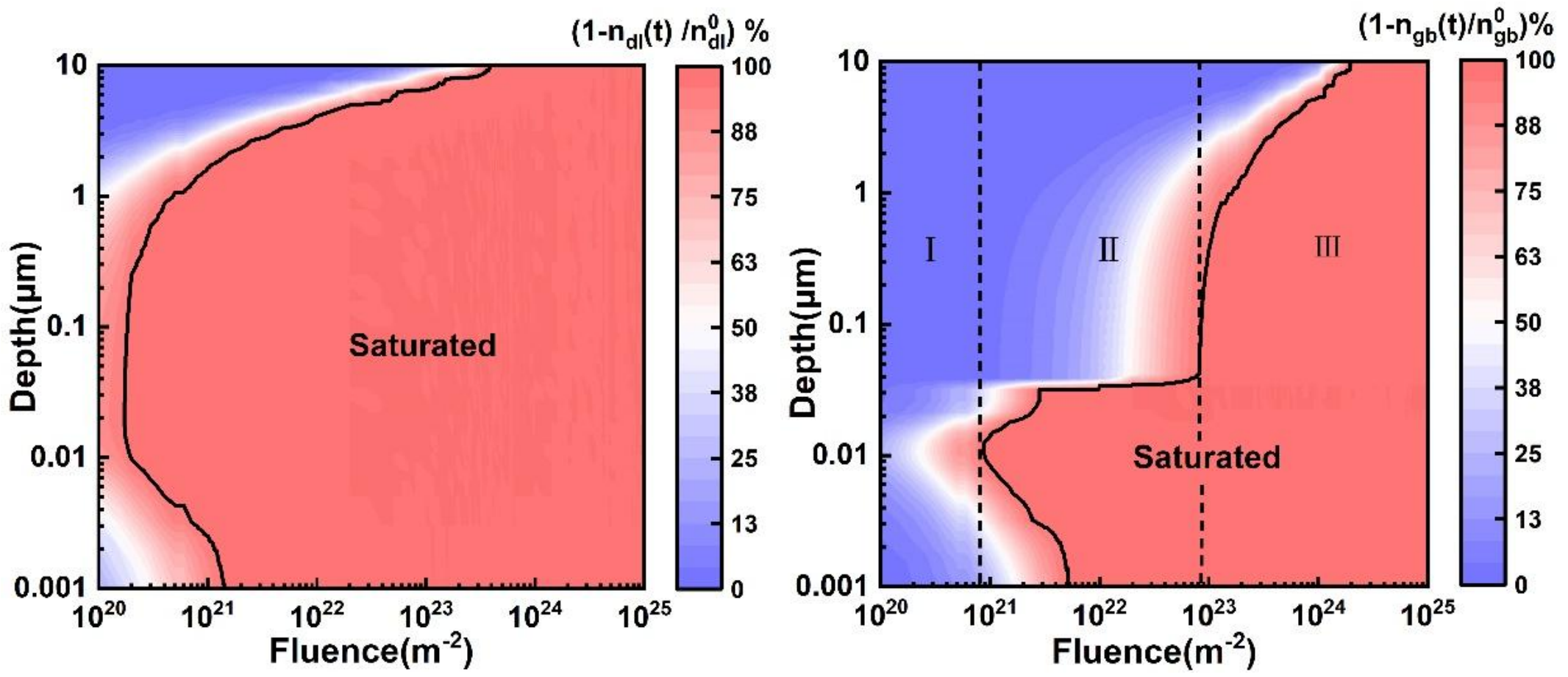

**Fig. 8** Distributions of occupied TIS density ratios of (a) DLs and (b) GBs with depth and fluence for 500 eV D ion implantation with the flux of $5\times10^{19}$ $m^{-2}\cdot s^{-1}$ at 300 K. Black lines represent 95% occupied TIS density ratios.

The trapping-site sink strength model quantifies D trapping behavior in inherent defects/microstructures like DLs and GBs by converting their structural characteristics into equivalent site densities. In addition, the concentration of ion-induced vacancies can also be converted into site concentrations, allowing D retention behavior in W to be evaluated within a unified framework. The total D trapping capacity of all these sinks is characterized by the ratio of the total retained D to the total concentration of available sites $C_D / C_S^0$. The black line in Fig. 9 represents $C_D / C_S^0 = 1$, which corresponds to the threshold of D saturation retention. When $C_D / C_S^0 > 1$, all trapping sites are fully occupied by D, indicating saturated D retention. When $C_D / C_S^0 < 1$, some defects/sinks remain available for further D trapping, corresponding to an unsaturated retention. In this study, GBs are found to contribute dominantly to D retention, accounting for 40-60% near the surface (< 0.02 μm) and up to 99% beyond 0.04 μm. Vacancy-type clusters contribute approximately 10-30% near the surface, but their contribution decreases with increasing fluence due to enhanced I-V recombination. In contrast, DLs contribute less than 0.1% to D retention across all depths, due to their low concentration of sites, which is always negligible. These results highlight the dominant role of GBs in governing low-energy saturation of D retention in W.

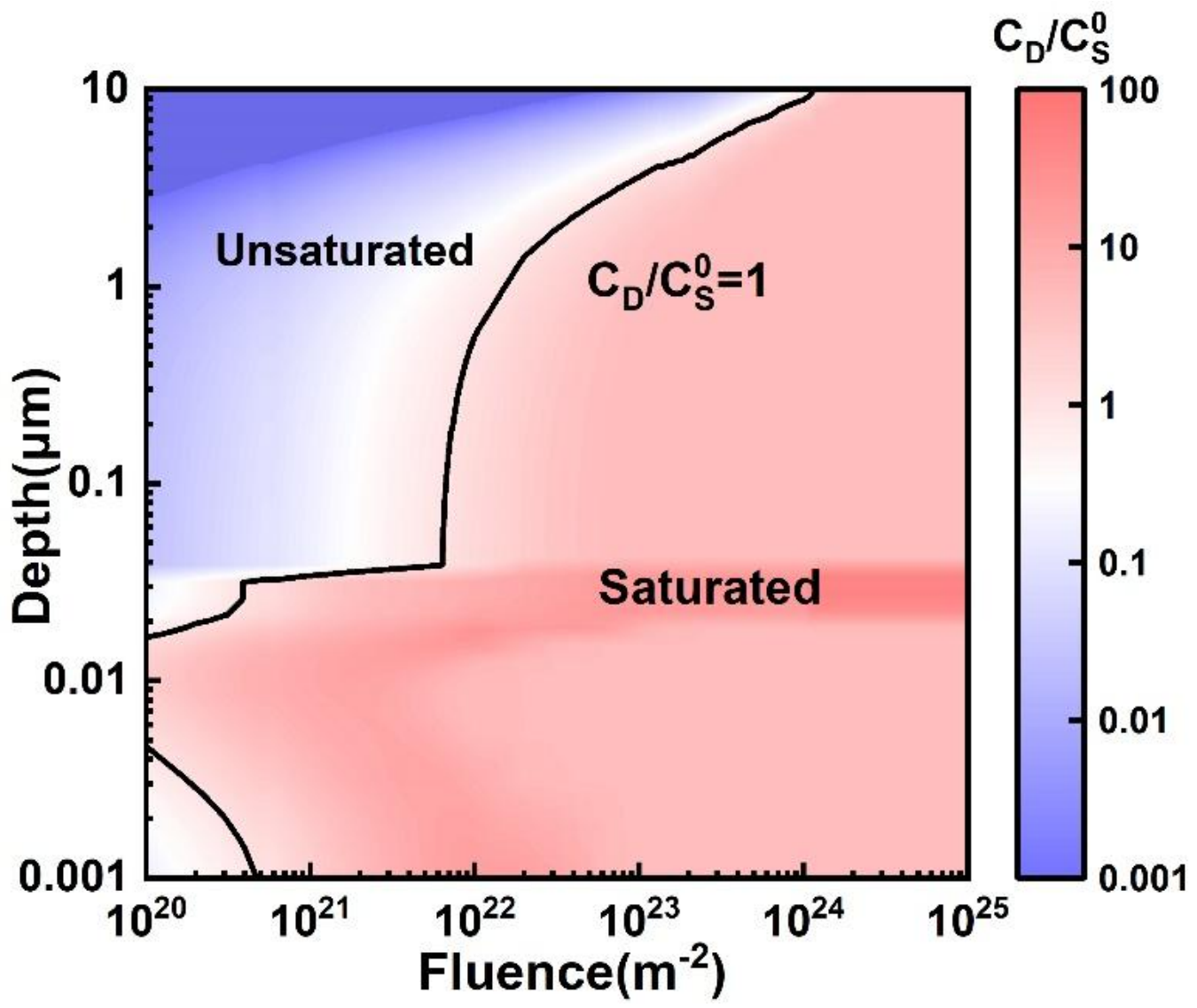

**Fig. 9** Distribution of $C_D / C_S^0$ in percent with depth and fluence for 500 eV D ion implantation with the flux of $5\times10^{19}$ $m^{-2}\cdot s^{-1}$ at 300 K. The black line represents $C_D / C_S^0=1$.

### 3.4 Basic prediction of D retention in microstructures

The trapping-site sink strength model offers a practical and unified framework to quantify the H trapping capacity of diverse microstructures. By converting their structural characteristics into effective site concentrations, this model enables a direct evaluation of their influence on HI retention in PFMs. In realistic PFMs applications, irradiation and material design can introduce extra traps beyond DLs and GBs, including voids, pre-irradiation bubbles, impurities, and second-phase particles (ODS/CDS) or solute atoms. To evaluate the contribution of various microstructures on D retention, common ranges of their structural parameters are selected for W-based PFMs reported in literatures, as given in Table 3.

Table 3 shows the site density on vacancies, impurity, DLs, GBs and ODS/CDS under typical irradiation conditions. This trapping-site model approximately treats 0-D sinks including vacancies/voids/cavities and impurity atoms as spherical trapping centers, whose sink strength is related to their respective concentrations and radii. As shown in Table 3, taking monovacancy as an example of vacancy-type defects, we estimated $n_{\mathrm{V}}^{0}$ is equal to 8, as widely given by atomic-scale simulations [48]. For larger-size voids/cavities, $n_{\mathrm{Void/Cavity}}^{0}$ is proportional to their sizes. Impurity atoms are assumed to occupy positions similar to vacancy and have a similar radius, whose site density $n_{\mathrm{Imputity}}^{0}$ is also approximately set as 8. Accordingly, the site density $n_{\mathrm{ODS/CDS}}^{0}$ on the interface of 2-D sinks in ODS/CDS is estimated about $10^{16}$ $m^{-2}$, by regarding the second-phase particles as an ensemble of impurity atoms. The site densities of 1-D (DL) and 2-D (GB) sinks are described in detail in Section 2.2. In Table 3, it shows that under typical conditions vacancies and ODS/CDS provide the highest site concentrations to dominate H trapping, followed by GBs and impurities, while DLs provide several orders of magnitude lower site concentrations, contributing the least to H retention.

Based on the trapping-site sink strength model, we can thus evaluate the influence of different types of sinks on H retention behavior under a unified framework. Under three typical irradiation conditions (H plasma irradiation alone, and its synergistic irradiation with neutrons or heavy ions), H retention in polycrystalline W arises from contributions of different types of sinks (especially for vacancy clusters and GBs). For low-energy H ion irradiation, H-induced vacancies are mainly distributed in the near-surface region on the nanometer scale, contributing little to the total H retention. Therefore, H retention primarily originates from trapping at GBs. For the synergistic irradiation of low-energy H ions and neutrons, neutron irradiation generates a large number of uniformly distributed vacancy-type clusters in W. These vacancy clusters can trap large amounts of H, thereby significantly increasing H retention. In this case, the dominant contributions to H retention come from both irradiation-induced vacancy clusters and GBs. For the synergistic irradiation of low-energy H ions and heavy ions, heavy ions introduce numerous vacancy-type clusters within a micrometer-scale depth range [65], leading to high H retention in the H implantation region. Meanwhile, the rapidly diffusing H atoms can also be trapped at deeper locations mainly by GBs, forming a spatially partitioned retention mechanism.

**Table 3** Site densities in diverse sinks with typical characteristics.

| Type | Sink strengths Formula | Typical range | Sink density | Theoretical sink strengths | Site density | $C_S^0$ |
|---|---|---|---|---|---|---|
| V | $4\pi r_V C_V$ [66] | $10^{-3}$-2 dpa | $10^{25}$-$10^{28}$ m$^{-3}$ | $10^{16}$-$10^{20}$ m$^{-2}$ | 8 | $10^{25}$-$10^{29}$ m$^{-3}$ |
| Impurities | $4\pi r_{Imputity} C_{Imputity}$ [66] | 0.1-200 appm | $10^{22}$-$10^{26}$ m$^{-3}$ | $10^{15}$-$10^{17}$ m$^{-2}$ | 8 | $10^{22}$-$10^{27}$ m$^{-3}$ |
| DLs | (5) | $10^{10}$-$10^{14}$ m$^{-2}$ | $10^{10}$-$10^{14}$ m$^{-2}$ | $10^{10}$-$10^{14}$ m$^{-2}$ | $10^{10}$ m$^{-1}$ | $10^{20}$-$10^{24}$ m$^{-3}$ |
| GBs | (6) | 0.1-100 μm | $10^{4}$-$10^{7}$ m$^{-1}$ | $10^{9}$-$10^{15}$ m$^{-2}$ | $10^{20}$ m$^{-2}$ | $10^{24}$-$10^{27}$ m$^{-3}$ |
| ODS | $4\pi r_{ODS} C_{ODS}$ [67] | 1-5 wt% | $10^{8}$-$10^{10}$ m$^{-1}$ | $10^{17}$-$10^{18}$ m$^{-2}$ | $10^{19}$ m$^{-2}$ | $10^{27}$-$10^{29}$ m$^{-3}$ |

Ultimately, the behavior of HI retention and radiation damage in W-based PFMs under realistic conditions can be basically predicted, providing guidance for the optimal regulation of sink density. At low temperatures below 700 K that the migration of vacancies is not activated, there exists a significant bias effect between SIAs and vacancies at sinks [68], which means that a higher sink density leads to poorer radiation damage resistance. Moreover, a high sink density also increases H retention. Therefore, in low-temperature regions such as the first-wall, materials with a low sink density should be selected to simultaneously reduce radiation damage and H/He retention [30]. At high temperatures above 700 K, the bias effect between SIAs and vacancies at sinks can be neglected [69], and increasing the sink density effectively enhances the absorption of intrinsic defects, thereby improving radiation damage resistance. However, this competes with the tendency of higher sink densities to increase H retention. Therefore, in high-temperature regions such as the divertor, an optimal sink density should be carefully selected to maximize radiation damage resistance while minimizing H retention.

## 4. Conclusion

A universal trapping-site sink strength model has been developed to quantitatively describe HI retention behavior in W, replacing conventional various analytical sink strength with an equivalent site concentration framework. For DLs and GBs, the model accurately reproduces the saturation and depth profiles of D retention under low-energy implantation. Distinct spatial retention mechanisms are revealed as follows: in irradiated regions, D trapping is governed by the combined effects of GBs and ion-induced vacancies; in the bulk, GBs dominate D trapping as D diffuses inward. The site concentration framework provides a unified metric for evaluating sink strength across various microstructures. The contributions to total trapping of GBs and vacancies are competitive in near-surface irradiated regions, GBs account for up to 99% in the bulk, whereas DLs play a negligible role (<0.1%) across all depths. The model quantitatively reveals that H retention originates from irradiation-induced vacancies and sinks through site concentration analysis, proposing tailored inhibition strategies for different irradiation conditions. This approach provides an efficient tool for optimizing the irradiation performance of PFMs in fusion reactors, by enabling basic predictions of HI retention behavior through a simplified assessment of the trapping capacities of various microstructures, rather than relying on large-scale simulations.

## Acknowledgements

This work was supported by the National Natural Science Foundation of China (Grant No. 12375277), the Strategic Priority Research Program of the Chinese Academy of Sciences (Grant No. XDA0410000), the Anhui Provincial Natural Science Foundation (Grant No. 2308085J04), and the National Foreign Experts Project (Type Y) by the Chinese Government (Grant No. Y20240239). Our calculations were performed at the Center for Computational Science of CASHIPS, the ScGrid of Supercomputing Center and the Computer Network Information Center of the Chinese Academy of Sciences. Numerical computations were also performed at Hefei advanced computing center.

## References

[1] Wang T, Ren M, Zhu X-L, Ma X, Yuan Y, Cheng L and Lu G-H 2022 Effect of initial exposure temperature on the deuterium retention and surface blistering in tungsten *Nucl. Mater. Energy* **33** 101245

[2] Kato D, Iwakiri H, Watanabe Y, Morishita K and Muroga T 2015 Super-saturated hydrogen effects on radiation damages in tungsten under the high-flux divertor plasma irradiation *Nucl. Fusion* **55** 083019

[3] Persianova A P and Golubeva A V 2024 Hydrogen Traps in Tungsten: A Review *Phys. Met. Metallogr.* **125** 278–306

[4] Causey R A 2002 Hydrogen isotope retention and recycling in fusion reactor plasma-facing components *J. Nucl. Mater.*

[5] Marchhart T, Hargrove C, Marin A, Schamis H, Saefan A, Lang E, Wang X and Allain J P 2024 Discovering tungsten-based composites as plasma facing materials for future high-duty cycle nuclear fusion reactors *Sci. Rep.* **14** 13864

[6] Li Y-G, Zheng Q-R, Wei L-M, Zhang C-G and Zeng Z 2020 A review of surface damage/microstructures and their effects on hydrogen/helium retention in tungsten *Tungsten* **2** 34–71

[7] Bakaev A, Grigorev P, Terentyev D, Bakaeva A, Zhurkin E E and Mastrikov Yu A 2017 Trapping of hydrogen and helium at dislocations in tungsten: an *ab initio* study *Nucl. Fusion* **57** 126040

[8] Song C, Hou J, Chen L, Liu C S and Kong X-S 2024 Bridging the gap between theory and experiment in vacancy concentration, C/N/O diffusivity, and divacancy interaction in tungsten: Role of vacancy-C/N/O interaction *Acta Mater.* **263** 119516

[9] Huang L, Jiang L, Topping T D, Dai C, Wang X, Carpenter R, Haines C and Schoenung J M 2017 In situ oxide dispersion strengthened tungsten alloys with high compressive strength and high strain-to-failure *Acta Mater.* **122** 19–31

[10] El-Atwani O, Cunningham W S, Esquivel E, Li M, Trelewicz J R, Uberuaga B P and Maloy S A 2019 In-situ irradiation tolerance investigation of high strength ultrafine tungsten-titanium carbide alloy *Acta Mater.* **164** 547–59

[11] Chen Y-S, Lu H, Liang J, Rosenthal A, Liu H, Sneddon G, McCarroll I, Zhao Z, Li W, Guo A and Cairney J M 2020 Observation of hydrogen trapping at dislocations, grain boundaries, and precipitates *Science* **367** 171–5

[12] Dubinko V I, Grigorev P, Bakaev A, Terentyev D, Van Oost G, Gao F, Van Neck D and Zhurkin E E 2014 Dislocation mechanism of deuterium retention in tungsten under plasma implantation *J. Phys. Condens. Matter* **26** 395001

[13] Chen H, Wang L, Peng F, Xu Q, Xiong Y, Zhao S, Tokunaga K, Wu Z, Ma Y, Chen P, Luo L and Wu Y 2023 Hydrogen retention and affecting factors in rolled tungsten: Thermal desorption spectra and molecular dynamics simulations *Int. J. Hydrog. Energy* **48** 30522–31

[14] Markelj S, Zavašnik J, Šestan A, Schwarz-Selinger T, Kelemen M, Punzón-Quijorna E, Alberti G, Passoni M and Dellasega D 2024 Deuterium retention and transport in ion-irradiated tungsten exposed to deuterium atoms: Role of grain

boundaries *Nucl. Mater. Energy* **38** 101589
[15] Zheng X-R, Kong X-S, Liu C S and Zhou X 2025 The role of grain boundary character on hydrogen energetics and kinetics in tungsten: Insights from atomic-scale modeling *Acta Mater.* 121267
[16] Wang S, Zhang J, Luo L-M, Zan X, Xu Q, Zhu X-Y, Tokunaga K and Wu Y-C 2016 Properties of Lu2O3 doped tungsten and thermal shock performance *Powder Technol.* **301** 65–9
[17] Jiang X, Sergienko G, Kreter A, Brezinsek S and Linsmeier Ch 2021 In situ study of short-term retention of deuterium in tungsten during and after plasma exposure in PSI-2 *Nucl. Fusion* **61** 096006
[18] Wang P, Cao Q, Hou J, Kong X-S, Chen L and Xie Z M 2022 Implantation and desorption of H isotopes in W revisited by object kinetic Monte Carlo simulation *J. Nucl. Mater.* **561** 153576
[19] Malerba L, Becquart C S and Domain C 2007 Object kinetic Monte Carlo study of sink strengths *J. Nucl. Mater.* **360** 159–69
[20] Jansson V, Malerba L, De Backer A, Becquart C S and Domain C 2013 Sink strength calculations of dislocations and loops using OKMC *J. Nucl. Mater.* **442** 218–26
[21] Valles G, Panizo-Laiz M, González C, Martin-Bragado I, González-Arrabal R, Gordillo N, Iglesias R, Guerrero C L, Perlado J M and Rivera A 2017 Influence of grain boundaries on the radiation-induced defects and hydrogen in nanostructured and coarse-grained tungsten *Acta Mater.* **122** 277–86
[22] Brailsford A D, Bullough R and Hayns M R 1976 Point defect sink strengths and void-swelling *J. Nucl. Mater.* **60** 246–56
[23] Bullough R, Hayns M R and Wood M H 1980 Sink strengths for thin film surfaces and grain boundaries *J. Nucl. Mater.* **90** 44–59
[24] Wood M H 1983 Sink strengths for grain-boundary cavities *J. Nucl. Mater.* **119** 67–72
[25] Bullough R and Quigley T M 1981 Dislocation sink strengths for the rate theory of irradiation damage *J. Nucl. Mater.* **104** 1397–401
[26] Jin X, Djurabekova F, Hodille E A, Markelj S and Nordlund K 2024 Analysis of lattice locations of deuterium in tungsten and its application for predicting deuterium trapping conditions *Phys. Rev. Mater.* **8** 043604
[27] Mathew N, Perez D and Martinez E 2020 Atomistic simulations of helium, hydrogen, and self-interstitial diffusion inside dislocation cores in tungsten *Nucl. Fusion* **60** 026013
[28] Xiao W and Geng W T 2012 Role of grain boundary and dislocation loop in H blistering in W: A density functional theory assessment *J. Nucl. Mater.* **430** 132–6
[29] Ning R H, Li Y G, Zhou W H, Zeng Z and Ju X 2012 Modeling D retention in W under D ions and neutrons irradiation *J. Nucl. Mater.* **430** 20–6
[30] Zhao Z, Li Y, Zhang C, Pan G, Tang P and Zeng Z 2017 Effect of grain size on the behavior of hydrogen/helium retention in tungsten: a cluster dynamics modeling *Nucl. Fusion* **57** 086020

[31] Li Y G, Zhou W H, Huang L F, Zeng Z and Ju X 2012 Cluster dynamics modeling of accumulation and diffusion of helium in neutron irradiated tungsten *J. Nucl. Mater.* **431** 26–32
[32] Chen X, Zhang Y, Wei L, Zheng Q, Zhang C and Li Y 2024 Cluster dynamics modeling of hydrogen retention and desorption in tungsten with saturation and multi-trapping effect of sinks *Nucl. Fusion* **64** 096037
[33] Cheng F, Li Y, Zheng Q, Wei L, Zhang C, Da B and Zeng Z 2023 Sensitivity of ion implantation to low-energy electronic stopping cross-sections *Radiat. Phys. Chem.* **204** 110681
[34] Alimov V K, Shu W M, Roth J, Sugiyama K, Lindig S, Balden M, Isobe K and Yamanishi T 2009 Surface morphology and deuterium retention in tungsten exposed to low-energy, high flux pure and helium-seeded deuterium plasmas *Phys. Scr.* **T138** 014048
[35] Jiang Z, Xia T, Han W, Shi Y, Zhang W and Zhu K 2024 Lateral stress induced blistering of tungsten exposed to deuterium plasma *Phys. Scr.* **99** 105310
[36] Ning R H, Li Y G, Zhou W H, Zeng Z and Ju X 2012 AN IMPROVED CLUSTER DYNAMICS MODEL FOR HYDROGEN RETENTION IN TUNGSTEN *Int. J. Mod. Phys. C* **23** 1250042
[37] Pokor C, Brechet Y, Dubuisson P, Massoud J-P and Barbu A 2004 Irradiation damage in 304 and 316 stainless steels: experimental investigation and modeling. Part I: Evolution of the microstructure *J. Nucl. Mater.* **326** 19–29
[38] Christien F and Barbu A 2009 Cluster Dynamics modelling of irradiation growth of zirconium single crystals *J. Nucl. Mater.* **393** 153–61
[39] Hardouin Duparc A, Moingeon C, Smetniansky-de-Grande N and Barbu A 2002 Microstructure modelling of ferritic alloys under high flux 1 MeV electron irradiations *J. Nucl. Mater.* **302** 143–55
[40] Donghua Xu, Gerrit VanCoevering and Brian D. Wirth 2016 Defect microstructural equivalence in molybdenum under different irradiation conditions at low temperatures and low doses *Comput. Mater. Sci.* **114**
[41] Ahlgren T, Heinola K, Vörtler K and Keinonen J 2012 Simulation of irradiation induced deuterium trapping in tungsten *J. Nucl. Mater.* **427** 152–61
[42] Jourdan T, Bencteux G and Adjanor G 2014 Efficient simulation of kinetics of radiation induced defects: A cluster dynamics approach *J. Nucl. Mater.* **444** 298–313
[43] Nagata S and Takahiro K 2000 Deuterium retention in tungsten and molybdenum *J. Nucl. Mater.* **283–287** 1038–42
[44] Kong X-S, Wang S, Wu X, You Y-W, Liu C S, Fang Q F, Chen J-L and Luo G-N 2015 First-principles calculations of hydrogen solution and diffusion in tungsten: Temperature and defect-trapping effects *Acta Mater.* **84** 426–35
[45] Liu Y-L, Zhang Y, Luo G-N and Lu G-H 2009 Structure, stability and diffusion of hydrogen in tungsten: A first-principles study *J. Nucl. Mater.* **390–391** 1032–4
[46] Henriksson K O E, Nordlund K, Krasheninnikov A and Keinonen J 2005 Difference in formation of hydrogen and helium clusters in tungsten *Appl. Phys. Lett.* **87** 163113

[47] Sun L, Liu Y-N, Xiao W and Zhou M 2018 Hydrogen behaviors at the near-surface region of tungsten: A first-principles study *Mater. Today Commun.* **17** 511–6

[48] Hou J, Kong X-S, Wu X, Song J and Liu C S 2019 Predictive model of hydrogen trapping and bubbling in nanovoids in bcc metals *Nat. Mater.* **18** 833–9

[49] Alimov V K and Roth J 2007 Hydrogen isotope retention in plasma-facing materials: review of recent experimental results *Phys. Scr.* **T128** 6–13

[50] Li Y G, Zhou W H, Ning R H, Huang L F, Zeng Z and Ju X 2012 A Cluster Dynamics Model For Accumulation Of Helium In Tungsten Under Helium Ions And Neutron Irradiation *Commun. Comput. Phys.* **11** 1547–68

[51] Mannheim A, Van Dommelen J A W and Geers M G D 2018 Modelling recrystallization and grain growth of tungsten induced by neutron displacement defects *Mech. Mater.* **123** 43–58

[52] Ning R H, Li Y G, Zhou W H, Zeng Z and Ju X 2012 Modeling D retention in W under D ions and neutrons irradiation *J. Nucl. Mater.* **430** 20–6

[53] Oude Vrielink M A, Shah V, Van Dommelen J A W and Geers M G D 2021 Modelling the brittle-to-ductile transition of high-purity tungsten under neutron irradiation *J. Nucl. Mater.* **554** 153068

[54] Yang X and Oyeniyi W O 2017 Kinetic Monte Carlo simulation of hydrogen diffusion in tungsten *Fusion Eng. Des.* **114** 113–7

[55] Becquart C S, Domain C, Sarkar U, DeBacker A and Hou M 2010 Microstructural evolution of irradiated tungsten: Ab initio parameterisation of an OKMC model *J. Nucl. Mater.* **403** 75–88

[56] Olsson P A T 2009 Semi-empirical atomistic study of point defect properties in BCC transition metals *Comput. Mater. Sci.* **47** 135–45

[57] Ohsawa K, Toyama T, Hatano Y, Yamaguchi M and Watanabe H 2019 Stable structure of hydrogen atoms trapped in tungsten divacancy *J. Nucl. Mater.* **527** 151825

[58] Grigorev P, Bakaev A, Terentyev D, Van Oost G, Noterdaeme J-M and Zhurkin E E 2017 Interaction of hydrogen and helium with nanometric dislocation loops in tungsten assessed by atomistic calculations *Nucl. Instrum. Methods Phys. Res. Sect. B Beam Interact. Mater. At.* **393** 164–8

[59] J.P. Roszell, A.A. Haasz and J.W. Davis 2011 D retention in W due to 10–500eV D + irradiation *J. Nucl. Mater.* **415**

[60] Ogorodnikova O V, Roth J and Mayer M 2003 Deuterium retention in tungsten in dependence of the surface conditions *J. Nucl. Mater.* **313–316** 469–77

[61] Wright G M, Mayer M, Ertl K, De Saint-Aubin G and Rapp J 2010 Hydrogenic retention in irradiated tungsten exposed to high-flux plasma *Nucl. Fusion* **50** 075006

[62] Alimov V Kh, Roth J and Mayer M 2005 Depth distribution of deuterium in single- and polycrystalline tungsten up to depths of several micrometers *J. Nucl. Mater.* **337–339** 619–23

[63] Ogorodnikova O V, Tyburska B, Alimov V Kh and Ertl K 2011 The influence of radiation damage on the plasma-induced deuterium retention in self-implanted

tungsten *J. Nucl. Mater.* **415** S661–6
[64] Ma P-W, Mason D R, Van Boxel S and Dudarev S L 2024 Nanocrystalline tungsten at high radiation exposure *Phys. Rev. Mater.* **8** 083601
[65] Wang S, Guo W, Schwarz-Selinger T, Yuan Y, Ge L, Cheng L, Zhang X, Cao X, Fu E and Lu G-H 2023 Dynamic equilibrium of displacement damage defects in heavy-ion irradiated tungsten *Acta Mater.* **244** 118578
[66] Ahlgren T and Bukonte L 2017 Sink strength simulations using the Monte Carlo method: Applied to spherical traps *J. Nucl. Mater.* **496** 66–76
[67] Dellasega D, Alberti G, Fortuna-Zalesna E, Zielinski W, Pezzoli A, Möller S, Unterberg B, Passoni M and Hakola A 2023 Nanostructure formation and D retention in redeposited-like W exposed to linear plasmas *Nucl. Mater. Energy* **36** 101492
[68] Wei L, Li Y, Zhao G, Zheng Q, Li J and Zeng Z 2019 Key factors in radiation tolerance of BCC metals under steady state *Nucl. Instrum. Methods Phys. Res. Sect. B Beam Interact. Mater. At.* **455** 134–9
[69] Wei L, Zhao Z, Li Y, Zheng Q, Zhang C, Li J, Zhao G, Da B and Zeng Z 2024 Absorption bias: A descriptor for radiation tolerance of polycrystalline BCC metals *J. Nucl. Mater.* **600** 155295